\documentclass[twoside,twocolumn,9pt]{article}
\usepackage{amsmath}
\usepackage{amsfonts}
\usepackage{amssymb}
\usepackage{mathrsfs}
\usepackage{extsizes}
\usepackage[super,sort&compress,comma]{natbib} 
\usepackage[version=3]{mhchem}
\usepackage[left=1.5cm, right=1.5cm, top=1.785cm, bottom=2.0cm]{geometry}
\usepackage{balance}
\usepackage{times,mathptmx}
\usepackage{sectsty}
\usepackage{graphicx} 
\usepackage{lastpage}
\usepackage[format=plain,justification=justified,singlelinecheck=false,font={stretch=1.125,small,sf},labelfont=bf,labelsep=space]{caption}
\usepackage{float}
\usepackage{fancyhdr}
\usepackage{fnpos}
\usepackage[english]{babel}
\usepackage{array}
\usepackage{droidsans}
\usepackage{charter}
\usepackage[T1]{fontenc}
\usepackage[usenames,dvipsnames]{xcolor}
\usepackage{setspace}
\usepackage[compact]{titlesec}
\usepackage{hyperref}

\usepackage{soul}
\numberwithin{equation}{section} \numberwithin{figure}{section}
\numberwithin{table}{section}

\newcommand{\be}{\begin{equation}}
\newcommand{\ee}{\end{equation}}

\newcommand{\bse}{\begin{subequations}}
\newcommand{\ese}{\end{subequations}}
\def\benl{\begin{eqnarray*}}
\def\eenl{\end{eqnarray*}}

\newcommand{\ben}{\begin{eqnarray}}
\newcommand{\een}{\end{eqnarray}}
\newcommand{\beq}{\begin{equation}}
\newcommand{\eeq}{\end{equation}}
\newcommand{\bea}{\begin{aligned}}
\newcommand{\eea}{\end{aligned}}
\newcommand{\bef}{\begin{figure}[H]}
\newcommand{\eef}{\end{figure}}

\definecolor{cream}{RGB}{222,217,201}
\begin{document}
\pagestyle{fancy}
\thispagestyle{plain}
\fancypagestyle{plain}{

%%%HEADER%%%
\fancyhead[C]{\includegraphics[width=18.5cm]{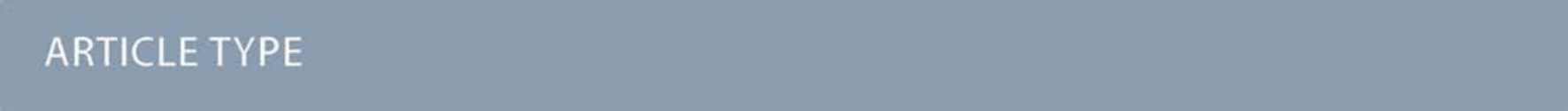}}
\fancyhead[L]{\hspace{0cm}\vspace{1.5cm}\includegraphics[height=30pt]{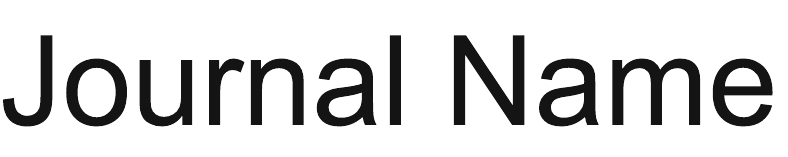}}
\fancyhead[R]{\hspace{0cm}\vspace{1.7cm}\includegraphics[height=55pt]{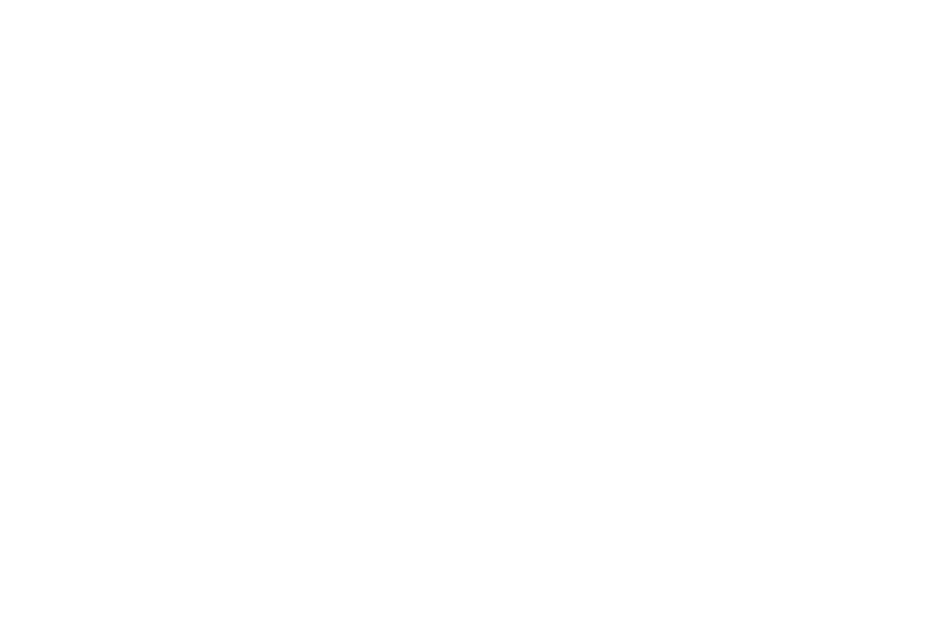}}
\renewcommand{\headrulewidth}{0pt}
}
%%%END OF HEADER%%%

%%%PAGE SETUP - Please do not change any commands within this section%%%
\makeFNbottom
\makeatletter
\renewcommand\LARGE{\@setfontsize\LARGE{15pt}{17}}
\renewcommand\Large{\@setfontsize\Large{12pt}{14}}
\renewcommand\large{\@setfontsize\large{10pt}{12}}
\renewcommand\footnotesize{\@setfontsize\footnotesize{7pt}{10}}
\makeatother

\renewcommand{\thefootnote}{\fnsymbol{footnote}}
\renewcommand\footnoterule{\vspace*{1pt}% 
\color{cream}\hrule width 3.5in height 0.4pt \color{black}\vspace*{5pt}} 
\setcounter{secnumdepth}{5}

\makeatletter 
\renewcommand\@biblabel[1]{#1}            
\renewcommand\@makefntext[1]% 
{\noindent\makebox[0pt][r]{\@thefnmark\,}#1}
\makeatother 
\renewcommand{\figurename}{\small{Fig.}~}
\sectionfont{\sffamily\Large}
\subsectionfont{\normalsize}
\subsubsectionfont{\bf}
\setstretch{1.125} %In particular, please do not alter this line.
\setlength{\skip\footins}{0.8cm}
\setlength{\footnotesep}{0.25cm}
\setlength{\jot}{10pt}
\titlespacing*{\section}{0pt}{4pt}{4pt}
\titlespacing*{\subsection}{0pt}{15pt}{1pt}
%%%END OF PAGE SETUP%%%

%%%FOOTER%%%
\fancyfoot{}
\fancyfoot[LO,RE]{\vspace{-7.1pt}\includegraphics[height=9pt]{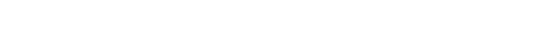}}
\fancyfoot[CO]{\vspace{-7.1pt}\hspace{13.2cm}\includegraphics{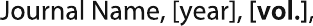}}
\fancyfoot[CE]{\vspace{-7.2pt}\hspace{-14.2cm}\includegraphics{head_foot/RF}}
\fancyfoot[RO]{\footnotesize{\sffamily{1--\pageref{LastPage} ~\textbar  \hspace{2pt}\thepage}}}
\fancyfoot[LE]{\footnotesize{\sffamily{\thepage~\textbar\hspace{3.45cm} 1--\pageref{LastPage}}}}
\fancyhead{}
\renewcommand{\headrulewidth}{0pt} 
\renewcommand{\footrulewidth}{0pt}
\setlength{\arrayrulewidth}{1pt}
\setlength{\columnsep}{6.5mm}
\setlength\bibsep{1pt}
%%%END OF FOOTER%%%

%%%FIGURE SETUP - please do not change any commands within this section%%%
\makeatletter 
\newlength{\figrulesep} 
\setlength{\figrulesep}{0.5\textfloatsep} 

\newcommand{\topfigrule}{\vspace*{-1pt}% 
\noindent{\color{cream}\rule[-\figrulesep]{\columnwidth}{1.5pt}} }

\newcommand{\botfigrule}{\vspace*{-2pt}% 
\noindent{\color{cream}\rule[\figrulesep]{\columnwidth}{1.5pt}} }

\newcommand{\dblfigrule}{\vspace*{-1pt}% 
\noindent{\color{cream}\rule[-\figrulesep]{\textwidth}{1.5pt}} }

\makeatother
%%%END OF FIGURE SETUP%%%

%%%TITLE, AUTHORS AND ABSTRACT%%%
\twocolumn[
  \begin{@twocolumnfalse}
\vspace{3cm}
\sffamily
\begin{tabular}{m{4.5cm} p{13.5cm} }

\includegraphics{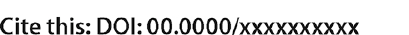} & \noindent\LARGE{\textbf{Solution landscapes of the diblock copolymer-homopolymer model under two-dimensional confinement}} \\%Article title goes here instead of the text "This is the title"
\vspace{0.3cm} & \vspace{0.3cm} \\

  & \noindent\large{Zhen Xu,\textit{$^{a}$} Yucen Han,\textit{$^{a, b}$} Jianyuan Yin,\textit{$^{c}$} Bing Yu,\textit{$^{c}$} Yasumasa Nishiura,$^{\ast}$\textit{$^{d}$} and Lei Zhang$^{\ast}$\textit{$^{a, e}$}} \\%Author names go here instead of "Full name", etc.

\includegraphics{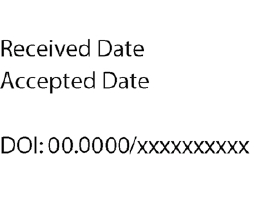} & \noindent\normalsize{
We investigate the solution landscapes of the confined diblock copolymer and homopolymer in two-dimensional domain by using the extended Ohta--Kawasaki model.
The projected saddle dynamics method is developed to compute the saddle points with mass conservation and construct the solution landscape by coupling with downward/upward search algorithms.
A variety of novel stationary solutions are identified and classified in the solution landscape, including \textit{Flower} class, \textit{Mosaic} class, \textit{Core-shell} class, and \textit{Tai-chi} class. 
The relationships between different stable states are shown by either transition pathways connected by index-$1$ saddle points or dynamical pathways connected by a high-index saddle point. 
The solution landscapes also demonstrate the symmetry-breaking phenomena, in which more solutions with high symmetry are found when the domain size increases.
%This 2D solution landscape study is a kind of compass for experimentalists and laid a foundation for further investigation of more complex multi-phases in three dimension.
} \\
\end{tabular}

 \end{@twocolumnfalse} \vspace{0.6cm}

  ]
%%%END OF TITLE, AUTHORS AND ABSTRACT%%%

%%%FONT SETUP - please do not change any commands within this section
\renewcommand*\rmdefault{bch}\normalfont\upshape
\rmfamily
\section*{}
\vspace{-1cm}

%%%FOOTNOTES%%%

\footnotetext{\textit{$^{a}$ Beijing International Center for Mathematical Research, Peking University, Beijing 100871, China.}}% E-mail: xuzhen@bicmr.pku.edu.cn}}
\footnotetext{\textit{$^{b}$ Department of Mathematics and Statistics, University of Strathclyde, G1 1XQ, UK.}}%E-mail: 
\footnotetext{\textit{$^{c}$ School of Mathematical Sciences, Peking University, Beijing 100871, China. }}%E-mail: 
\footnotetext{\textit{$^{d}$ Research Institute for Electronic Science, Hokkaido University, N12W7, Kita-Ward, Mid-Campus Open Laboratory Building No.2, Sapporo 060-0812, Japan, MathAM-OIL, Tohoku University and AIST, Japan;  Email: yasumasa@pp.iij4u.or.jp}}
\footnotetext{\textit{$^{e}$ Center for Quantitative Biology, Peking University, Beijing 100871, China. E-mail: zhangl@math.pku.edu.cn}}

%%%END OF FOOTNOTES%%%

%%%MAIN TEXT%%%%
\section{Introduction}
The diblock copolymers are composed of two different blocks linked together through covalent bonding. %not monomers
A large number of copolymers interact with each other to produce a wide variety of microstructures, resulting from a compromise between phase segregation and polymer architecture.~\cite{Bates1996,Hamley1998,Bates2000,DengLiuLTZ2012} 
The diblock copolymer materials have aroused great interest in both industrial and theoretical research.
Moreover, an important commercialized application of copolymers is thermoplastic elastomers, which have been widely used as jelly candles, outer coverings for optical fiber cables, adhesives, bitumen modifiers, etc.~\cite{Bhowmick2001,Craver2000,Holden2001}
Much scientific interest on the self-assembly of block copolymers is due to the pattern formation and the potential applications of the microstructures led by the confinement mechanism, which restricts degrees of freedom in space and breaks symmetry of the structure.~\cite{Park2003Poly}

Extensive experimental and theoretical studies have demonstrated that confinement can be used to control the self-assembly of diblock copolymers.~\cite{ShiLi2013Rev} 
The diblock copolymers under different confinements have been well studied, such as two-dimensional (2D) cylindrical confinement,~\cite{HePan2001JCP,SevinkHuinink2001,Xiang2004Macro,LiWickGar2006} 3D cylindrical confinement,~\cite{Shin2007,Feng2007JCP,Dobriyal2009Macro}
and 3D spherical or polyhedral confinement.~\cite{YuLiShi2007Macro,YangShi2012sphere,Li2013sphere} 
Many novel morphologies have been discovered, ~\cite{Wu2004,Xiang2004Macro,Thomas2008,ShiLi2013Rev} 
for example, the onion-like and layered structures for symmetric copolymers under spherical confinement in {experiments}.~\cite{Nishiura2016,Nishiura2018}
These resulting morphologies are often distinctly different from those in the bulk phase~\cite{Nishiura2010} and thus defined as the "frustrated phases".~\cite{Yabu2014SoftRev}
Meanwhile, many mathematical models and numerical studies have been carried out to investigate the confinement of copolymers and their self-assembly, including Monte Carlo simulation,~\cite{HePan2001JCP,YuLiShi2007Macro,Feng2007JCP,HanJiang2008Macro} cell dynamics simulation,~\cite{PinnaZve2008Poly, PinnaZve2009JCP, PinnaZve2010ACS,DengShi2015Macro} the self-consistent field theory,~\cite{ChenShi2007Macro,GuoShi2008PRL,XuJiangShi2013JPC} and the phase field method.~\cite{OhtaIto1995,Choksi2005Phys,Gennip2008PDE,Glasner2019JAM}

% model
%%% Ohta-Kawasaki: homopolymer-copolymer and micro,macro-sepatation 
%The Ohta-Kawasaki model, derived by the mean field theory, is a phase field model to describe the system composed of the block copolymer or/and homopolymer.~\cite{OhtaKawasaki1986,Nishiura1995,Nishiura2010,Nishiura2016,Nishiura2018} The free-energy functional of the Ohta-Kawasaki model is in the form of a generalized Landau free energy functional with nonlocal terms to describe the linkage between the different blocks in copolymer.
%%governing the formation and structure of the confined copolymer
%Two phase variables are introduced to characterize the macro-phase separation between the homopolymer and copolymer, as well as the micro-phase separation between the two components of the diblock copolymers.~\cite{Nishiura1995, Ito1998}
%%DCH:
The first qualitative model for the block copolymers was proposed by Ohta and Kawasaki~\cite{OhtaKawasaki1986} in the form of a generalized Landau free-energy functional with nonlocal term to describe the linkage between the different blocks in copolymers. This was carefully re-derived in a more mathematical formulation by Nishiura et al.~\cite{Nishiura1995} and its singular limit was also discussed there.  With this introduction to mathematical community, Choksi and many other people then started to work in this direction.~\cite{Choksi2005Phys,Nishiura2010} The above Ohta-Kawasaki model was extended to a system composed of block copolymers and homopolymers with two phase variables to describe the macro-phase separation between the copolymers and homopolymers, as well as the micro-phase separation between the two components of the diblock copolymers.~\cite{OhtaIto1995,Ito1998} Hereafter we call the extended Ohta--Kawasaki free-energy functional as the diblock copolymer-homopolymer (DCH) free-energy functional for simplicity.
Multiple stable self-assembled phases can be found by the minimization of the DCH functional, such as layers, tennis balls, onions, and multipods under the nano-confinements.~\cite{YuLiShi2007Macro,YangShi2012sphere,Li2013sphere, Nishiura2016}
However, the relationships between different stable states and the landscape of possible stationary solutions of the DCH free-energy functional are still unclear.

% method
The stationary solutions correspond to the solutions of the Euler-Lagrange equation of the DCH free-energy functional with the mass conservation. 
The Euler-Lagrange equation usually has multiple solutions, including both stable/metastable solutions, i.e., local minima, and unstable saddle points of the system.
%These stationary solutions can be distinguished by the Morse index of the solution.
The properties of  stationary solutions can be characterized by the Morse index of the solution.
The Morse index of a stationary solution is equal to the number of negative eigenvalues of its Hessian matrix.~\cite{Milnor1963} In particular, a local minimum or stable state can be regarded as an index-$0$ solution with no unstable directions. 
Compared to computing a stable state by gradient dynamics, the saddle point is much more difficult to find due to its unstable nature, while often plays critical roles in determining the properties of the model system.
For instance, to find the transition pathways between two stable states, one needs to compute the transition state, which is an index-$1$ saddle point and the corresponding Hessian matrix has one and only one negative eigenvalue.
It has attracted substantial attentions to find multiple stationary solutions of the nonlinear problems. \cite{zhang2016recent} Considerable efforts have been made to develop various numerical algorithms, such as the minimax method,~\cite{LiZhou2001} the deflation technique,~\cite{Farrell2015} the eigenvector-following method,~\cite{Doye2002} and the homotopy method.~\cite{Mehta2011,Hao2014} In particular, Yin et al. proposed a novel saddle dynamics (SD) and implemented a high-index optimization-based shrinking dimer method to compute any-index saddle points.~\cite{zhang2016optimization, YinZhangZhangSISC2019} By combining the SD with the downward and upward search algorithms, Yin et al. further constructed a solution landscape, which is a pathway map consisting of all stationary solutions and their connections, for the unconstrained systems.\cite{YinWangZhangPRL2020} 
This numerical approach has been successfully applied to the Landau--type free-energy functional, including the defect landscape of confined nematic liquid crystal on a square using a Landau--de Gennes model~\cite{YinWangZhangPRL2020, HanMajumdarZhang2020} and the transition pathways between period crystals and quasicrystals by applying the Lifshitz--Petrich model.~\cite{YinJiangShiZhang2020}

%what we do
In this paper, we apply the DCH model to investigate the solution landscape of the diblock copolymers and homopolymers in 2D confinement.
To deal with the mass-conservation constraint, we develop the projected saddle dynamics (PSD) method, which is a constrained version of the SD by using the projection operator.
Then we systematically construct the solution landscapes with two critical parameters: one represents the preference intensity and the other corresponds to the domain size.
A variety of novel stationary solutions are found in the solution landscapes, including \textit{Flower} class, \textit{Mosaic} class, \textit{Core-shell} class, and \textit{Tai-chi} class. 
Furthermore, the solution landscapes demonstrate the relationships between different stable states by either transition pathways connected by index-1 saddle points or dynamical pathways connected by a high-index saddle point. 
The solution landscapes also reveal the symmetry-breaking phenomena, in which more solutions with high symmetry are identified when the domain size increases.

% % outline
The rest of the paper is organized as follows. 
The DCH model is briefly introduced in Section 2.
The PSD method and the numerical algorithm of construction of a solution landscape are presented in Section 3.
 We numerically construct the solution landscapes with different preferences and domain sizes in Section 4.
 Final conclusions and discussions are presented in Section 5.
 
\section{Diblock copolymer-homopolymer model}
%Now we first give a brief description for the DCH model, who has a Landau free energy functional of dibolock copolymers.
%This phase field model was first considered by Ohta and Kawasaki,~\cite{OhtaKawasaki1986} and was then carefully re-derived in a more mathematical formulation.~\cite{Nishiura1995} With this introduction to mathematical community, Choksi and many other people then started to work in this direction.~\cite{Choksi2005Phys}
%%The Ohta-Kawasaki (OK) model is a famous phase field model with one component to describe the diblock copolymer.
We consider the mixture of AB diblock copolymers and C homopolymers,~\cite{OhtaIto1995} with two independent and conserved phase-field order parameters $\eta$ and $\phi$. 
$\eta$ represents the macro-phase separation with a phase boundary that can be understood as a confining surface, which arises naturally to separate the homopolymer phase from the copolymer phase. %$-1$ corresponds to a homopolymer rich domain and  $+1$ corresponds to a copolymer rich domain.  $\eta\in[-1,1]$
The copolymers are assumed to be immersed in an external-medium homopolymers or solvent, such as water. 
In the copolymer-rich domain, another variable $\phi$ describes the micro-phase separation between the block A and block B. %, taking values from the interval $[-1,1]$. The end points corresponds to A and B, respectively. 
 When the above two systems interact with one another, the morphologies consist of the confinement surface and copolymer components within the surface, and then undergo a macro-phase and micro-phase separation described by $\eta$ and $\phi$, respectively.

%The dynamics of the two mixed systems evolves to minimize the value of an free-energy functional in the following Landau expression:
The DCH free-energy functional can be written as a sum of short-range contribution and long-range contribution
\begin{equation}\label{eq:energy}
F\{\eta;\phi\}=F_S\{\eta;\phi\}+F_L\{\eta;\phi\}.
\end{equation}
%namely, which can be written as a sum of short-range and long-range contributions.
The short-range contribution $F_S$ is given by
\begin{equation}
F_S\{\eta;\phi\}=\int_{\Omega} \left[\frac{D_1}{2}|\nabla\eta|^2+\frac{D_2}{2}|\nabla\phi|^2+W(\eta,\phi)\right]d\textbf{r},
\end{equation}
where $\Omega$ is a Lipschitz-boundary domain in $\mathbb{R}^2$.
%We mainly works on 2D confinement, hence $\Omega$ is a smooth bounded domain in $\mathbb{R}^2$. The model can also be studied in 3D domain. 
$D_1, D_2$ are parameters controlling the size of the macro-phase and micro-phase separation interface, respectively. %interface between macrophases and microphases. 

The potential is taken as the polynomial form, %Nishiura:double-well/triple-well
%\begin{eqnarray}
%W(\eta,\phi)&\!=\!&\frac{(\eta^2-1)^2}{4}+\frac{(\phi^2-1)^2}{4}  \nonumber \\
%&& +b_1\eta\phi-\frac{b_2}{2}\eta\phi^2-\frac{b_3}{2}\eta^2\phi+\frac{b_4}{2}\eta^2\phi^2. \label{W}
%\end{eqnarray}
\ben
\bea
W(\eta,\phi)=&\frac{(\eta^2-1)^2}{4}+\frac{(\phi^2-1)^2}{4}  \\
                     & +b_1\eta\phi-\frac{b_2}{2}\eta\phi^2-\frac{b_3}{2}\eta^2\phi+\frac{b_4}{2}\eta^2\phi^2. \label{W}
\eea
\een
The first two terms in \eqref{W} exhibit double-well potential for $\eta$ and $\phi$, respectively, and the rest terms describe the coupling between the AB copolymers and the solvent C.~\cite{OhtaIto1995, Ito1998}
The coefficients $b_1$, $b_2$, $b_3$ and $b_4$ are positive constants, which are related to the molecular parameters and could be derived in principle by the generalized method.~\cite{OhtaKawasaki1986,OhtaIto1995, Ito1998,HanXuShiZhang2020}
These parameters are chosen so that $W(\eta,\phi)$ has a triple-well structure with three distinct minima corresponding to the phases of block A, block B and solvent C, respectively.
%Here, we select the same values as in the work of \cite{Nishiura2016,Nishiura2018}, for example, $b_3=b_4=0$, $b_2=1$, $b_1\in[0,0.2]$.  
%With positive $\eta$, i.e., copolymer rich domain, 
Here, we set $b_3=b_4=0$ and note that $b_1, b_2$ alter slightly the minimum points of $W(\eta,\phi)$ from the ideal values of $(\pm 1,\pm 1)$.~\cite{Nishiura2016}
When $b_1 = 0$, the free-energy functional is symmetrical corresponding to $\phi$, indicating that $\eta$ or the confining surface has equal preference for positive or negative $\phi$, such as the morphology of layer.~\cite{Nishiura2016,Nishiura2018,HanXuShiZhang2020} 
Conversely, the nonzero $b_1$ would cause symmetry breaking between micro-phase separated domains, i.e., the selective preference between block A ($\phi>0$) and block B ($\phi<0$). 
%The interaction between the confined copolymer and the confining surface can also be controlled via tuning its value, such as the simulation of the onion morphology.~\cite{Nishiura2016,Nishiura2018}

The long-range contribution $F_L$ is given by
\begin{equation}
F_L\{\eta;\phi\}=\int_{\Omega}\frac{\alpha}{2}|(-\Delta)^{-\frac{1}{2}}(\phi-\bar{\phi})|^2d\textbf{r},
\end{equation}
where $\bar{\phi}$ represents the spatial average of $\phi$.
In the original paper, the Green functions are used to represent long-range interactions,~\cite{OhtaKawasaki1986} which was replaced with a non-local operator, the fractional power of the Laplace operator for variational problems.~\cite{Nishiura1995} 
The long-range contribution prevents the copolymers from forming a large macroscopic domain and brings about many fine structures, such as layers or onions.
$\alpha$ is inversely proportional to the square of the total chain length of the copolymer and related to the bonding between block A and block B in the copolymers,  hence is a measure of the connectivity between two blocks.~\cite{Choksi2009}
When $\alpha=0$, there is no linkage between A and B blocks, and the absence of the non-local term will induce the separation macroscopically. 
If $\alpha\neq 0$, we have microphases within the copolymer-rich domain and multiple morphologies arise. 
%We note that in the original formulation of the free-energy functional for the DCH model, the long-range contribution can be written in the following Green functions~\cite{OhtaKawasaki1986,OhtaIto1995}
%\ben
%\bea
%F_L\{\eta,\phi\}&=\int d\textbf{r}\int d\textbf{r}'G(r,r')\big[\frac{\alpha}{2}\delta \phi(r)\delta\phi(r')+\beta\delta\phi(r)\delta\eta(r') \\
%                       &+\frac{\gamma}{2}\delta\eta(r)\delta\phi(r')\big].
%\eea
%\een
%The Green's function $G(r,r')$ is defined by the relation $-\Grad^2G(r,r')=\delta(r-r')$, the field variables $\delta\eta=\eta-\bar{\eta}, \delta\phi=\phi-\bar{\phi}$ with $\bar{\eta}, \bar{\phi}$ representing the spatial average of $\eta$ and $\phi$, respectively. The coefficients $\alpha, \beta$ and $\gamma$ are related to the molecular parameters of the model system.
%For simplicity, we take $\beta=\gamma=0$, and $\alpha$ as the positive coefficient in the domain $(40, 200)$.

%Considering the effect of the domain size $\lambda$ for the 2D case, 
Now we nondimensionalize the system with $\tilde{r}=r/\lambda$, then the rescaled free-energy functional is 
\ben
\bea
\tilde{F}\{\tilde\eta;\tilde\phi\}&=\int_{\Omega}\frac{D_1}{2}|\tilde{\nabla}\tilde{\eta}|^2+\frac{D_2}{2}|\tilde{\nabla}\tilde{\phi}|^2+\lambda^2\left(\frac{1}{4}(\tilde{\eta}^2-1)^2+\frac{1}{4}(\tilde{\phi}^2-1)^2\right.   \\
                                            &\left.+b_1\tilde{\eta}\tilde{\phi}-\frac{1}{2}b_2\tilde{\eta}\tilde{\phi}^2\right)+\frac{\alpha\lambda^4}{2}|(-\tilde{\Delta})^{-\frac{1}{2}}(\tilde{\phi}-\bar{\phi})|^2d\tilde{r},\label{eq:F_lambda}
\eea
\een
where $\tilde{\Omega}$ is a unit square $[0,1]\times[0,1]$, and $\lambda$ is the length of the square domain.
%As $\lambda$ increases, the nonlinearity and non localization of \eqref{eq:F_lambda} become stronger. 
%Thus, we will have a more complicated solution landscape with variety of new solutions and interesting connections between them. 
In what follows, we drop the tildes and all statements are in terms of the rescaled variables.
Here both $\eta$ and $\phi$ are the conserved order parameters satisfying
\begin{equation}
\int_{\Omega}(\eta-\bar{\eta})d\textbf{r}=0, ~~\int_{\Omega}(\phi-\bar{\phi})d\textbf{r}=0. \label{constrain}
\end{equation}

The stationary solutions of the DCH functional with mass  conservation are the solutions of the Euler-Lagrange equations as follows, 
\ben
\bea
\frac{\delta F(\eta,\phi)}{\delta\eta}-\xi_\eta     &= 0,  \label{CH:eta} \\
\frac{\delta F(\eta,\phi)}{\delta\phi}-\xi_\phi      &= 0. \label{CH:phi} \\
%\frac{\delta F(\eta,\phi)}{\delta\eta}-\lambda  \frac{\delta c}{\delta\eta}        &= 0,  \label{CH:eta} \\
%\frac{\delta F(\eta,\phi)}{\delta\phi}-\lambda  \frac{\delta c}{\delta\phi}         &= 0. \label{CH:phi} \\
%\frac{\delta F(\eta,\phi)}{\delta\eta}    &= -D_1\Delta\eta+\lambda^2W_\eta(\eta,\phi) = 0,  \label{CH:eta} \\
%\frac{\delta F(\eta,\phi)}{\delta\phi}    &= -D_2\Delta\phi+\lambda^2W_\phi(\eta,\phi)+\alpha\lambda^4(-\Delta)^{-1}(\phi-\bar{\phi}) = 0. \label{CH:phi} \\
\eea
\een
$\xi_\eta$, and $\xi_\phi$ are the Lagrangian multipliers to keep the mass conservation.

\section{Numerical method} \label{numerical}% \textcolor{red}{}
\subsection{Projected saddle dynamics method}
To find the multiple stationary solutions of the Euler-Lagrange equations \eqref{CH:phi} with the mass conservation \eqref{constrain}, 
we need to develop an efficient numerical algorithm to compute the saddle points with mass conservation.
% and the relationship between them, we need to construct the solution landscapes by combining the SD with the downward and upward search algorithms.~\cite{YinWangZhangPRL2020} 
%The original SD is designed for unconstrained problems, while in the confined diblock copolymer problem we have the mass-conservation constraints for both $\eta$ and $\phi$.
%With the high-index saddle points, we can derive the local minima solutions(index-0 saddle points) and their transition pathways via transition states(index-1 saddle points).
The original SD method is designed for unconstrained gradient systems \cite{YinZhangZhangSISC2019}. Recently, Huang et al. proposed a constrained high-index saddle dynamics method to compute the constrained saddle points and construct the solution landscape with equality constraints by using Riemannian gradient and Hessian \cite{HuangCHiSD2020}. %which needs much computational costs on the eigenvectors of Hessians. 
In the DCH model, since the mass conservation is only a linear constraint, we propose a simple PSD method to compute index-$k$ saddle points ($k$-saddles) with the mass-conservation constraint.
Here, the projection is defined as follows,
\begin{equation}
P(\xi) = \xi - \int_\Omega\xi d\textbf{r},~\xi\in L^2(\Omega).
\end{equation} %combined with the SD~\cite{YinZhangZhangSISC2019} to compute any-index saddle points with the mass conservation. 
Both gradient and Hessian of $F(\eta; \phi)$ are updated by the projected forms. 
%$\nabla F(\cdot)=\left[\frac{\delta F}{\delta \eta}, \frac{\delta F}{\delta \phi}\right]^T$ should be replaced by $P\nabla F(\cdot)=\left[P\frac{\delta F}{\delta \eta}, P\frac{\delta F}{\delta \phi}\right]^T$. %the inner product $<\cdot, \cdot>_{\mathcal{H}}$ is defined as $<P(\cdot), P(\cdot)>_{L^2(\Omega)}$.  %$(P_\eta\frac{\delta\eta}{\delta\phi}, P_\phi\frac{\delta\eta}{\delta\phi})$. $(P_\eta,P_\phi) = (\eta-\bar{\eta}, \phi-\bar{\phi})$.
In addition, to eliminate the unphysical directions, we translate the order parameters $\eta$ and $\phi$ to $\hat{\eta}=\eta-\bar{\eta}$ and $\hat{\phi}=\phi-\bar{\phi}$ so that  $\int_{\Omega}\hat{\eta}d\textbf{r}=0, \int_{\Omega}\hat{\phi}d\textbf{r}=0$. In the following, we also drop the hats and all statements are in terms of the translated variables.

The PSD for computing a mass-conserved $k$-saddle ($k$-PSD) is governed by the following dynamic equations 
\ben\label{psd}
\bea
\dot{\eta}&=-P\frac{\delta F(\eta;\phi)}{\delta \eta}+2\sum_{j=1}^k\left\langle P\frac{\delta F(\eta;\phi)}{\delta\eta}v_j+P\frac{\delta F(\eta;\phi)}{\delta \phi}w_j,v_j\right\rangle,  \label{HiSD:etaphi} \\
\dot{\phi}&=-P\frac{\delta F(\eta;\phi)}{\delta \phi}+2\sum_{j=1}^k\left\langle P\frac{\delta F(\eta;\phi)}{\delta\eta}v_j+P\frac{\delta F(\eta;\phi)}{\delta \phi}w_j,w_j\right\rangle.
%\dot{\phi}&=-\Big(I-2\sum_{j=1}^kv_jv_j^T\Big)\frac{\delta F(\eta,\phi)}{\delta \phi}(\phi), 
\eea
\een 
%\begin{tabular}{c}0 \\ 0\end{tabular}
%\left.\begin{array}{c}v_1 \\w_1\end{array}\right.
The equations \eqref{HiSD:etaphi} allow $[\eta;\phi]$ to move along an ascent direction on the subspace $\mathcal{V}=span\left([v_1;w_1],\cdots,[v_k;w_k]\right)$, and a descent direction on the subspace $\mathcal{V}^{\perp}$, the orthogonal complement space of $\mathcal{V}$. % and $\mathcal{W}$ respectively. %on the constrained surface. 
 $\left[v_i;w_i\right] (i = 1 \cdots k)$ are the orthonormal and unit eigenvectors, i.e., $\langle [v_i;w_i],[v_j;w_j]\rangle=\delta_{i,j}$, corresponding to the smallest $k$ eigenvalues $\lambda_1,\cdots,\lambda_k$ of the Hessian $\nabla^2 F(\eta; \phi)$.
They can be obtained via the constrained optimization problem and governed by the following equations
\ben
\bea
\!\left[\begin{array}{c} \dot{v_i} \\ \dot{w_i} \end{array} \!\right]
= & -\left(I-
\!\left[\begin{array}{c} v_i \\ w_i \end{array} \!\right]\!\left[\begin{array}{c} v_i \\ w_i \end{array} \!\right]^T\!\!\!\!+2\sum_{j=1}^{i-1}\!\left[\begin{array}{c} v_j \\ w_j \end{array} \!\right]\!\left[\begin{array}{c} v_j \\ w_j \end{array} \!\right]^T\!\!\right)  \\ 
   &\! \left(\begin{array}{c} P \\ P \end{array} \!\right)\!   \nabla^2F(\eta;\phi)\!\left[\begin{array}{c} v_i \\ w_i \end{array} \!\right],~~i=1,\cdots,k.
\eea
\een\label{v_w}
Here $I$ is the identity operator. The $k$-saddle $\left[\eta;\phi\right]$ and $k$ direction variables $\left[v_{i};w_{i}\right]$ are coupled. We note that $v_{i}$ and $w_{i}$ are also needed to apply the projection $P$ to keep mass conservation, namely $\int_\Omega v_id\textbf{r}=\int_\Omega w_id\textbf{r}=0$.

To avoid the direct calculation of Hessians, we approximate Hessian by using the central difference scheme for directional derivations on $k$ dimers centered at $\left[\eta; \phi\right]$.~\cite{zhang2016optimization, YinZhangZhangSISC2019} 
The $i$th dimer has a direction of $\left[v_i;w_i\right]$ with a small dimer length $2l$ and then $\nabla^2F(\eta;\phi)$ is approximated by
\begin{equation}
\nabla^2F(\eta; \phi)\!\left[\begin{array}{c}v_i \\ w_i \end{array} \!\right] \approx\frac{\nabla F\left(\!\left[\begin{array}{c}\eta+lv_i  \\ \phi+lw_i \end{array} \!\right]\right) -\nabla F\left(\!\left[\begin{array}{c}\eta-lv_i  \\ \phi-lw_i \end{array}\!\right]\right) }{2l}.
\end{equation}

We choose periodic boundary conditions and apply the Fourier spectral method for the space discretization on $\Omega$. The numerical simulations are performed on a 2D $128\times128$ grid which was verified to give well-resolved numerical results. 
%The periodic boundary condition on 2D will lead to $2$ zero eigenvalues of Hessian $\nabla^2F(\eta,\phi)$ on stationary solution. 
We use the explicit Euler scheme and Barzilai--Borwein gradient method to determine the step sizes for time discretization of \eqref{HiSD:etaphi}.~\cite{BB1988} %Euler explicit discretization
Furthermore, we apply the locally optimal block preconditioned conjugate gradient (LOBPCG) method~\cite{Knyazev2001} to compute the smallest $k$ eigenvalues and the corresponding eigenvectors of the Hessian. 

The initial condition is given as:
\begin{equation}
\!\left[\begin{array}{c}\eta(0) \\ \phi(0) \end{array} \!\right]=\!\left[\begin{array}{c} \eta^0 \\ \phi^0 \end{array} \!\right],\!\left[\begin{array}{c}v_i(0) \\ w_i(0) \end{array} \!\right]=\!\left[\begin{array}{c} v_i^0 \\ w_i^0\end{array} \!\right],~i=1,\cdots,k,
\end{equation}
where $\left[\eta^0; \phi^0\right]$ are zero mean, and $\left[v_1^0; w_1^0\right],\cdots,\left[v_k^0;w_k^0\right]$ are the unit orthogonal vectors with zero mean.
%The self-adjoint Hessian here can be obtained from a constrained optimization problem,
%\begin{equation}
%\min_{v_i\in \mathbb{R}^n} <D^2F(x), v_i>, ~~s.t. ~~{v_j,v_i}=\delta_{ij},~ j=1,2,\cdots,i.
%\end{equation}

\subsection{Algorithm for the solution landscape}
The solution landscape of the DCH model is constructed via two algorithms: {downward search} and {upward search}.\cite{YinWangZhangPRL2020}
%The downward search method enables us to search for all connected lower-index saddle points from a $m$-saddle, and upward search with a selected direction can help to find the higher-index saddles, which drives the entire search to navigate up and down on the energy landscape.~\cite{YinYuZhangSCM2020} 
The downward search algorithm enables us to efficiently search for all connected low-index saddles and minima from a high-index saddle, and the upward search algorithm aims to find the possible higher-index saddles. The details of two algorithms coupled with PSD method are as follows:

\textit{Downward search algorithm}: 
Assuming we have an $m$-saddle $\left[\eta^{\star}; \phi^{\star}\right]$ and the $m$ normalized vectors $\left[v_1^{\star};w_1^{\star}\right],\cdots, \left[v_m^{\star}; w_m^{\star}\right]$ corresponding to the $m$ negative eigenvalues of the Hessian matrix $\nabla^2F(\eta^{\star};\phi^{\star})$. We then apply the $(m-1)$-PSD \eqref{psd} to search the $(m-1)$-saddles by choosing $\left[\eta^{\star};\phi^{\star}\right]\pm\epsilon \left[v_m^{\star};w_m^{\star}\right]$ as an initial state and $\left[v_i^{\star};w_i^{\star}\right],\cdots,\left[v_{m-1}^{\star};w_{m-1}^{\star}\right]$ as initial unstable directions. Once a new $(m-1)$-saddle is obtained, we continue to apply the $(m-2)$-PSD to search the $(m-2)$-saddles.

By repeating the above procedure, we can establish a systematic search for all saddle points branched from this $m-$saddle as a parent and to construct a family tree that eventually connects to the local minima.

\textit{Upward search algorithm}: 
If the parent state (the highest-index saddle point) is unavailable beforehand or multiple parent states exist, one can conduct the upward search to find the high-index saddle points starting from a local minimum or a low-index saddle point.

Starting from an $m$-saddle $\left[\eta^{\star}; \phi^{\star}\right]$, we apply the $(m+1)$-PSD to search an $(m+1)$-saddle. The initial state is chosen as $\left[\eta^{\star};\phi^{\star}\right]\pm\epsilon \left[v_{m+1}^{\star};w_{m+1}^{\star}\right]$, and $\left[v_{1}^{\star}; w_{1}^{\star}\right], \cdots, \left[v_{m+1}^{\star}; w_{m+1}^{\star}\right]$ are taken as the initial ascent directions, in which $\left[v_{m+1}^{\star};w_{m+1}^{\star}\right]$ is the eigenvector corresponding to the smallest positive eigenvalue of its Hessian matrix.

Each downward search represents the relaxation of a pseudo-dynamics, the so-called dynamical pathway, starting from a high-index saddle point to a local minimum. By combining the downward search and upward search, we are able to systematically find possible stationary solutions and uncover the connectivity of the solution landscape.

\section{Results}
Now we present the numerical results for the solution landscapes of the diblock copolymers and homopolymers under 2D confinement.
To see the effect of the preference intensity $b_1$ and the domain size $\lambda$, we choose three cases: the equal preference, the selective preference, and the equal preference in a larger domain. 

\subsection{Solution landscape with equal preference}
\begin{figure}[hbt]
\centering
\includegraphics[width=0.45\textwidth,clip==]{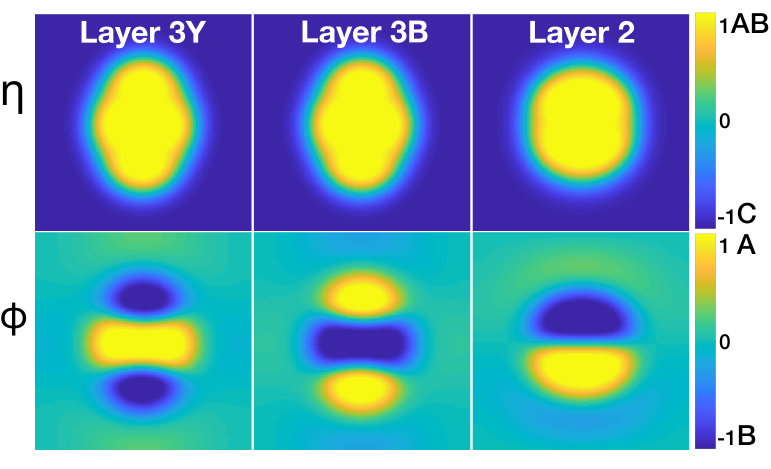}
\caption{Three stable states: $Layer~3Y$, $Layer~3B$ and $Layer~2$ with equal preference. The first row and the second row show the spatial distributions of $\eta$ and $\phi$, respectively.
In the plot of $\eta$, and AB copolymers and C homopolymers are represented by yellow and blue, respectively. 
In the plot of $\phi$, block A and block B of copolymers are distinguished by yellow and blue, respectively.
All subsequent figures share the same color bar for $\phi$.}\label{u0}
\end{figure}

%%  Fig. 1:Typical solution
In the case of equal preference ($b_1=0,~\lambda =1$), we plot three stable states ($Layer~3Y$, $Layer~3B$, and $Layer~2$) in Fig. \ref{u0}. Some parameters are set as $D_1=D_2=0.0025$, $b_2=1$, $\alpha=60$, $\bar{\eta}=-0.5$, $\bar{\phi}=0$. % all of which can be referred as the 2D cross section of the 3D Layers solution with lamellar structure and Janus solution with hemispheric block A and hemispheric block B confined in a sphere.~\cite{Nishiura2016} 
Since the preference for block A (yellow) and block B (blue) are equal, $Layer~3Y$ and $Layer~3B$ are a pair of solutions only with block A and B switched, and the two blocks in $Layer~2$ have the same area and shape. From an energy point of view, $Layer~2$ is the stable phase with the lowest energy, while $Layer~3Y$ and $Layer~3B$ are the metastable phases. 

\begin{figure}[hbt]
\centering
\includegraphics[width=0.4\textwidth,clip==]{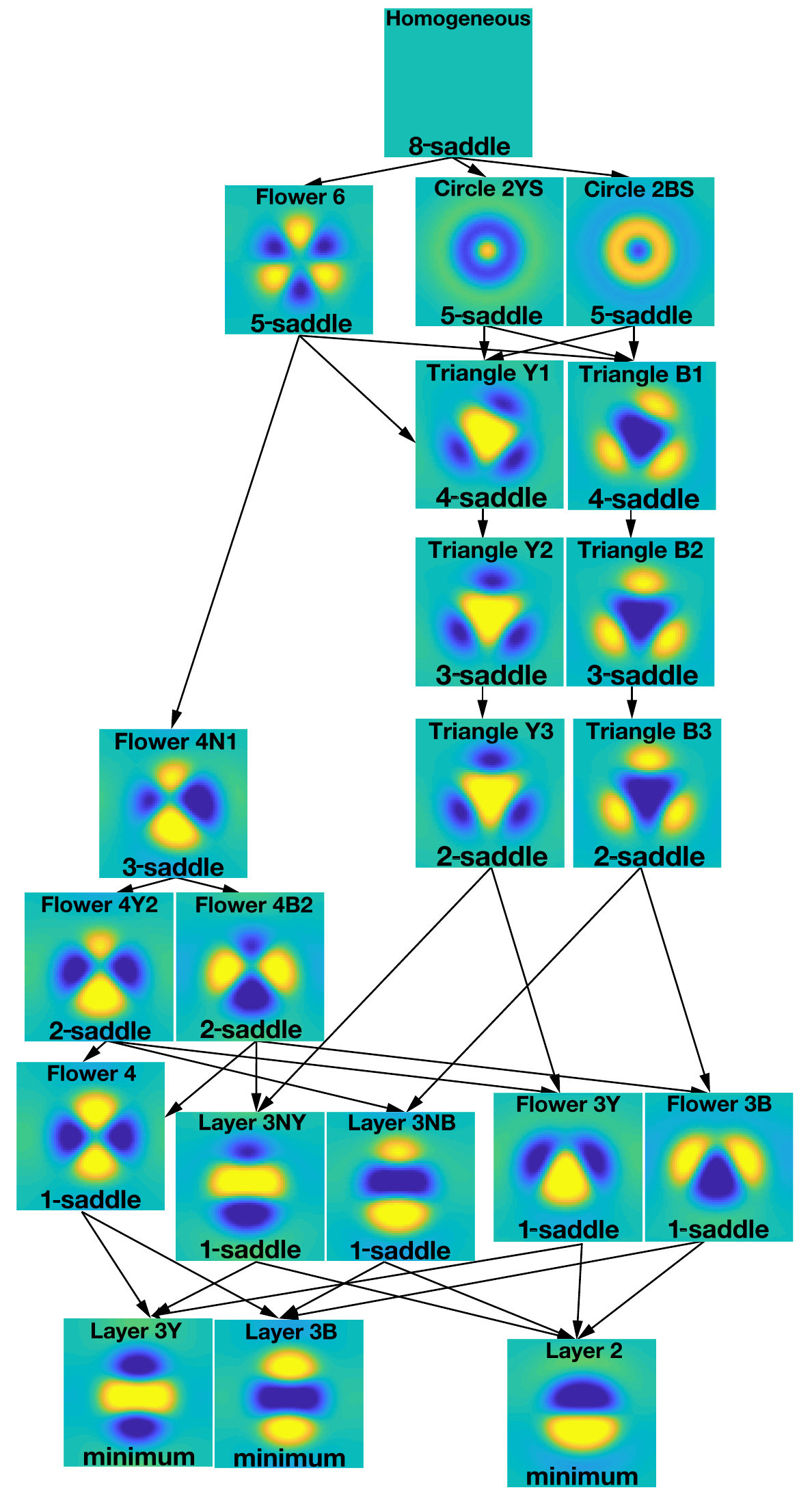}
\caption{Solution landscape with equal preference. The height of a phase approximately corresponds to its energy. The $\phi$ plot of each phase is shown in square domain with name on the top and index of corresponding saddle point at the bottom.} \label{sl_b0} 
\end{figure}

%%  Fig. 2: Solution landscape %
%In addition to these minima, the solution landscape can help us to find a global picture of stationary solutions, 
The solution landscape with equal preference is shown in Fig. \ref{sl_b0}. The homogeneous phase ($\eta\equiv\bar{\eta},\phi\equiv\bar{\phi}$) is clearly a trivial solution, which is a 8-saddle. Using it as the parent state, we are able to find three distinct 5-saddles via the downward search, specifically, $Flower~6$ looks like a blooming flower with three yellow petals alternating with three blue petals, and $Circle~2YS$ and $Circle~2BS$ are a pair of solutions due to the equal preference of blue blocks and yellow blocks. 

The periodic boundary condition implies that the Hessian at a non-homogeneous state has at least two zero eigenvalues in most cases, which explains why no 6-saddles or 7-saddle are found in Fig. \ref{sl_b0}. 
Down from these 5-saddles, a variety of complex morphologies in \textit{Triangle} class and \textit{Flower} class are obtained. In the \textit{Triangle} class, the shapes of inner blocks are isosceles triangles with the apex angle less than $60$ degrees ($Triangle~Y1$ and $Triangle~B1$), isosceles triangles with the apex angle greater than $60$ degrees ($Triangle~Y2$ and $Triangle~B2$), and equilateral triangles ($Triangle~Y3$ and $Triangle~B3$). In \textit{Flower} class, the asymmetric/symmetric flower solutions, $Flower~4N1$, $Flower~4Y2$, $Flower~4B2$, and $Flower~4$, have two yellow petals alternating with two blue petals. $Flower~3Y$ has one yellow petal between two blue petals and $Flower~3B$ has one blue petal between two yellow petals. 
The asymmetric $Layer~3NY$ ($Layer~3NB$) is the transition state between the metastable solution $Layer~3Y$ ($Layer~3B$) and the stable solution $Layer~2$ in Fig. \ref{u0}.

% % we find all the 0-saddles, namely the minimum solution down from all kinds of the 1-saddles.
% To make comparison with the solution landscape of other cases, we show both of the stationary solutions, considering the difference of block A and block B.
% Because of the periodic boundary condition, a translation of a nonhomogeneous stationary solution is also a stationary solution, so only one phase is shown in the solution landscape for simplicity. With the help of this kind of pathway map, these stationary solutions can be easily obtained and clearly illustrated, and we further find the transition pathways between the three local minimum solutions, shown in Fig. \ref{trans_b0}.
%%  Fig. 2: solution landscape

\begin{figure} [hbt]
\centering
\includegraphics[width=0.5\textwidth,clip==]{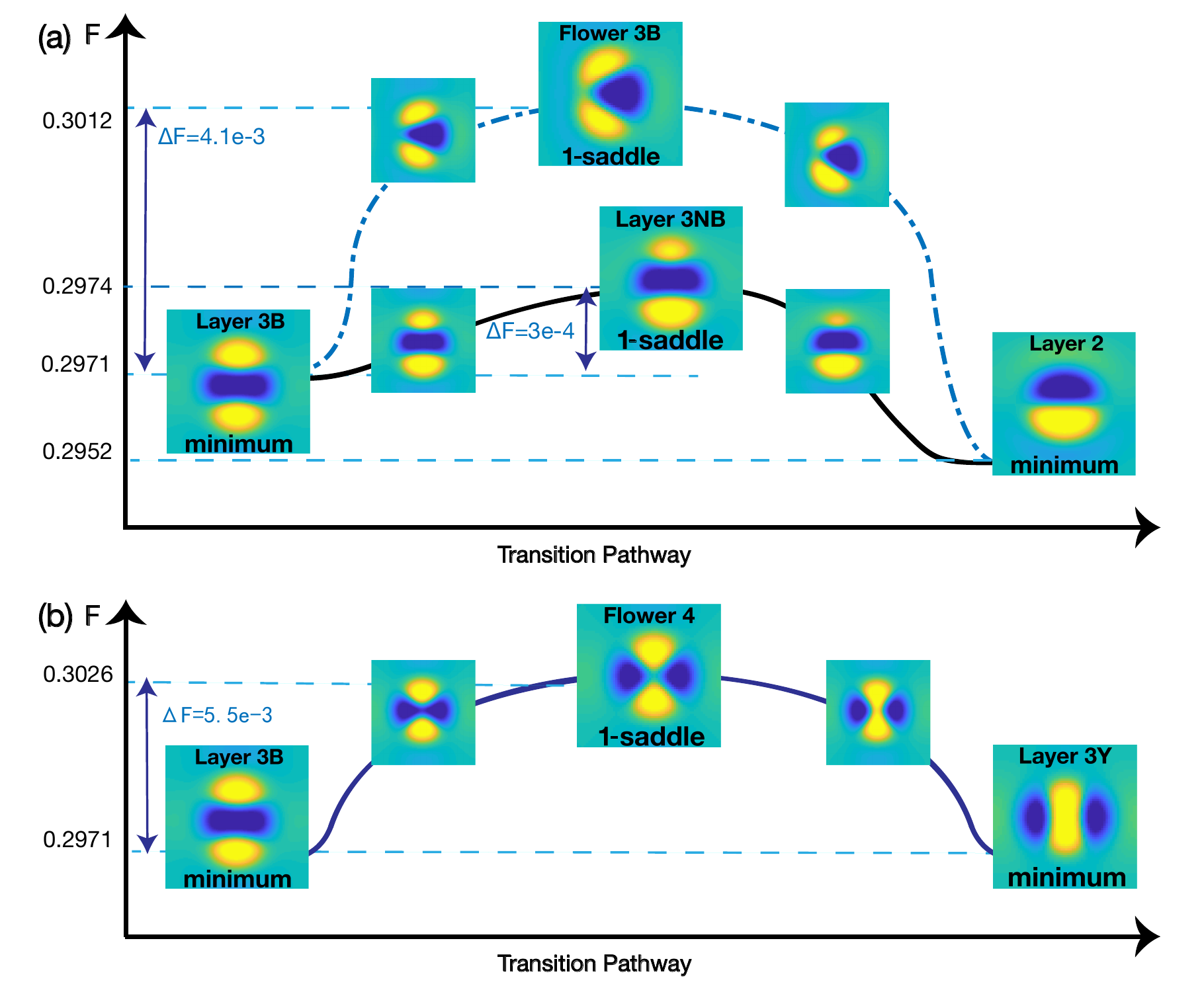}
\caption{Transition pathways between stable states with equal preference.
(a) Transition pathways between $Layer~3B$ and $Layer~2$ with the transition state $Flower\ 3B$ (energy barrier $4.1e-3$) or $Layer\ 3NB$ (energy barrier $3e-4$). (b) Transition pathway between $Layer~3B$ and $Layer~3Y$ with the transition state $Flower\ 4$ (energy barrier $5.5e-3$). }\label{trans_b0}
\end{figure}

%% Fig 3:  transition pathway
From the solution landscape, we can also extract the transition pathways between three stable states $Layer~3Y$, $Layer~3B$, and $Layer~2$ (Fig. \ref{trans_b0}). 
%The index-1 saddle which connects any two of the stable states along its unstable direction is the transition state along the transition pathway between them. 
There are two transition pathways between $Layer~3B$ and $Layer~2$: $Layer~3B \rightarrow Flower~3B \rightarrow Layer~2$ and $Layer~3B \rightarrow Layer~3NB \rightarrow Layer~2$ in Fig. \ref{trans_b0}(a). 
The latter one has the smaller energy barrier $\Delta F$ (the energy difference between the transition state and the initial stable state), thus has higher possibility to take place. With the equal preference, it is easy to see the transition pathways between $Layer~3Y$ and $Layer~2$ are analogous: $Layer~3Y \rightarrow Flower~3Y \rightarrow Layer~2$ ($\Delta F=4.1e-3$) and $Layer~3Y \rightarrow Layer~3NY \rightarrow Layer~2$ ($\Delta F=3e-4$).
%Also there are some other pathways with multiple transition states or higher-index saddle points to connect these stable states, while this is computationally expensive and need overcome higher energy barriers, e.g., the pathway $Layer~3Y \to Flower~4 \to Layer~3B \to Layer~3NB \to Layer~2$ or $Layer~3Y \to Layer~3NY \to Triangle~Y3 \to Flower~3B \to Layer~2$. These multiple unstable directions give us greater control on the dynamical pathways, all of which give greater insights into the design and control of solution landscapes. \cite{YinJiangShiZhang2020}. 
Fig. \ref{trans_b0}(b) shows the switching process between $Layer~3B$ and $Layer~3Y$, along which blue block and yellow block swap, connected by the transition state $Flower~4$ with two opposed blue petals and two opposed yellow petals. % process. %The transition pathway is $Layer~3B \rightarrow Flower~4 \rightarrow Layer~3Y$. 
In the process from $Layer~3B$ to $Flower~4$, the yellow block on both sides of the blue block push into the middle of blue block. The shape of blue block changes from an ellipse to an hourglass, and is further cut into two small petals. 
In the process from $Flower~4$ to $Layer~3Y$, the two yellow petals connect together resulting in a yellow hourglass and relax to an ellipse between two small blue ellipses. 
%The structural change between $Layer~3B$ and $Layer~3Y$ is localized in the central of the square domain.
From Fig. \ref{sl_b0}, $Layer~3B$ and  $Layer~3Y$ can also be connected via $Layer~2$, that is, $Layer~3B \rightarrow Layer~3NB \rightarrow Layer~2$ (Fig. \ref{trans_b0}(a)) and $Layer~2 \rightarrow Layer~3NY \rightarrow Layer~3Y$. In fact, this transition pathway has a lower energy barrier and is more probable than the one in Fig. \ref{trans_b0}(b).

%% Fig. 3:Transition pathway

\subsection{Solution landscape with selective preference}% different preference
\begin{figure}[hbt] 
\centering
\includegraphics[width=0.24\textwidth,clip==]{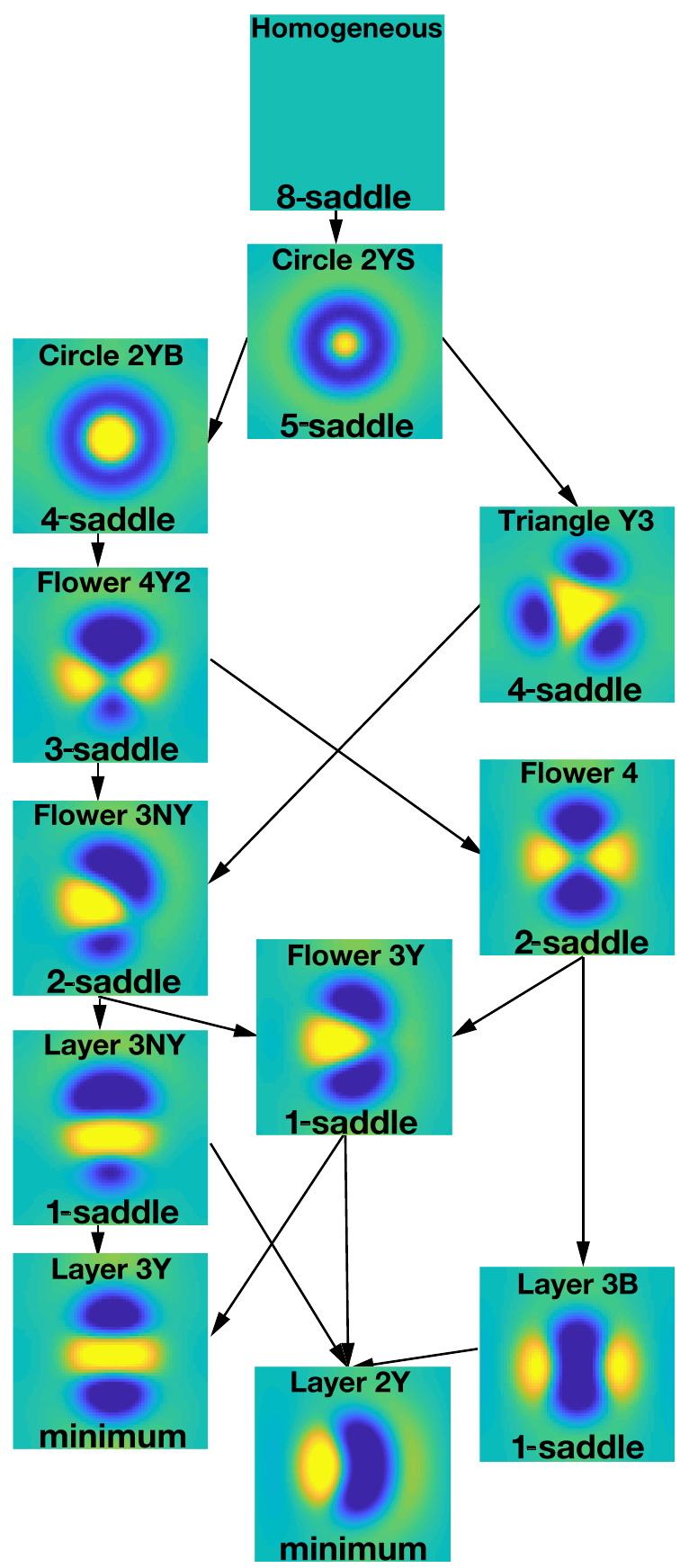}
\caption{Solution landscape with selective preference. The height of a phase approximately corresponds to its energy. The $\phi$ plot of each phase is shown in square domain with name on the top and index of corresponding stationary solution on the bottom.} \label{sl_b1} 
\end{figure}

 We next study the solution landscape with selective preference ($b_1=0.1,~\lambda=1$), namely the affinity of the blue block for homopolymers (solvent) is higher than that of the yellow one. This leads to the symmetry breaking of the morphologies in contrast to the one with equal preference. 
%This phenomena as $b_1$ varies are interesting to look at, as they shed light on how the parameters affect the behavior of a particular morphology.~\cite{Nishiura2016} 
In Fig. \ref{sl_b1}, the homogeneous phase (8-saddle) is still the parent state, and the transition states between $Layer~3Y$ and $Layer~2Y$ are also $Layer~3NY$ and $Flower~3Y$. 
%Block A has bigger area and more preference to the confining surface since $b1>0$. 
While due to the symmetry breaking, the structure of the solution landscape has changed dramatically. %and the structures of stationary solutions are similar to ones in Fig. \ref{sl_b0}. 
On one hand, we lose nearly half of the stationary solutions observed in Fig. \ref{sl_b0}. For instance, $Flower~6$ in Fig. \ref{sl_b0} with alternative yellow and blue blocks merges into $Circle~2YS$ with blue ring surrounding the yellow blocks at the center.
Although $Flower~6$ disappears in the case of selective preference, $Flower~4$ still survives with larger blue petals and smaller yellow petals.
Moreover, since the area of inner yellow blocks becomes smaller, $Flower~4N1$ in Fig. \ref{sl_b0} merges into $Flower~4Y2$, and $Triangle~Y1$ and $Triangle~Y2$ in Fig. \ref{sl_b0} merge into the regular triangle $Triangle~Y3$.
The $Layer~3B$ solution loses stability and becomes an $1$-saddle.
On the other hand, new solutions appear. For example, the $5$-saddle $Circle~2YS$ connects to a new $4$-saddle $Circle~2YB$, and the $1$-saddle $Flower~3Y$ bifurcates into a new asymmetric $2$-saddle $Flower~3NY$.

%In 3D simulation, when $b_1$ is big enough, block B is completely surrounded by block A and the onion phase will become a local minimum.~\cite{Nishiura2016} Hence, we suspect in our 2D work the same situation will take place. The index of \textit{Circle} class decrease as $b_1$ increases.

\subsection{Solution landscape with equal preference in a larger domain}
\begin{figure}[hbt] 
\centering
\includegraphics[width=0.4\textwidth,clip==]{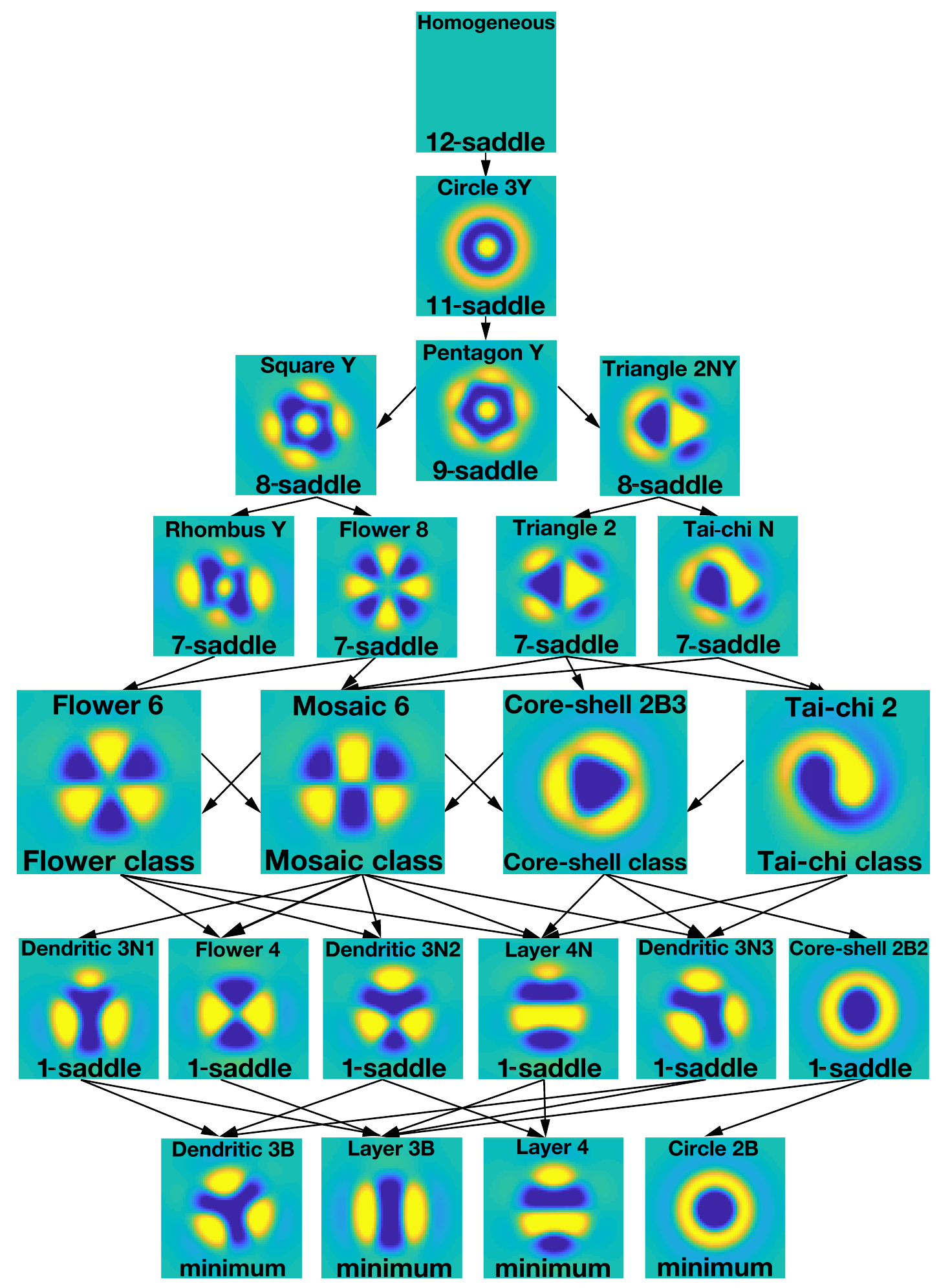}
\caption{Solution landscape with equal preference in a larger domain. The stationary solutions with index from $2$ to $6$ are classified by \textit{Flower} class, \textit{Mosaic} class, \textit{Core-shell} class and \textit{Tai-chi} class, and the typical solution is inserted in each class as an illustration.
The symmetric stationary solutions obtained by switching blue block and yellow one are omitted here.
%For instance, the $11$-saddle $Circle~3Y$ has a yellow core part, but blue one for core part ( $Circle~3B$) also exists as a $11$-saddle. 
}\label{sl_b14} 
\end{figure}

%We note that the domain size may not be consistent with the energy functional~\cite{Nishiura2016}. 
We further investigate the solution landscape with equal preference in a larger domain  ($b_1=0, \lambda=1.4$).
%In the rescaled free-energy functional \eqref{eq:F_lambda}, domain size $\lambda$ is the coefficient of the nonlinear term and nonlocal term. 
%As $\lambda$ increases, we expect more novel solutions and more complicated solution landscape.
In Fig. \ref{sl_b14}, the homogeneous state becomes a 12-saddle and more stationary solutions emerge in the solution landscape, such as 9-saddle $Pentagon~Y$ with the same symmetric property as a pentagon and 7-saddle $Flower~8$ with four blue petals and four yellow petals.
It is worth mentioning that we find novel stable solutions such as $Dendritic~3B$, $Layer~4$ and $Circle~2B$. 
%The solutions with exchanging block A and B in the $Dendritic~3$, $Layer~4$ and $Circle~2B$ are also stable since $b_1=0$. 
The $Dendritic~3B$ solution, which looks like a steering wheel, appears in the larger domain. This phase is bifurcated from $Triangle~Y3$ and is a 2D analogy to the multipod phase in 3D.~\cite{Nishiura2016}
The layer numbers of the \textit{Layer}-class solutions also increase to 3 (e.g. $Layer~3B$) and 4 (e.g. $Layer~4$). 
The $Circle~2B$ becomes a local minimum in the larger domain.

%% Fig 5: whole solution landscape 

To illustrate the solution landscape more clearly, we classify the stationary solutions from 2-saddles to 6-saddles into four classes: \textit{Flower} class, \textit{Mosaic} class, \textit{Core-shell} class, and \textit{Tai-chi} class, according to the configuration and connections between them.
\textit{Mosaic} indicates the morphology has multiple alternative yellow and blue blocks.
\textit{Core-shell} represents the class of morphologies that one of the block polymers is surrounded by the other one completely or partially.
It is similar to the \textit{Circle} class, but loses the symmetry of the circle. %The phases with symmetry $D^{\infty}$ are also classified by \textit{Circle} class.
\textit{Tai-chi} indicates a class of morphologies that both the blue and yellow parts seem like the $yin$ and $yang$ fishes engaged with each other.
If there exists one saddle point in \textit{Flower} class connecting to another saddle point in \textit{Mosaic} class, we present this connection by an arrow from \textit{Flower} to \textit{Mosaic} in Fig. \ref{sl_b14}. 
Hence the connections between four classes are \textit{Flower} $\leftrightarrow$ \textit{Mosaic} $\leftrightarrow$ \textit{Core-shell} $\leftarrow$ \textit{Tai-chi}, but the connection from \textit{Core-shell} class to \textit{Tai-chi} class has not been found yet. 
%$Flower$ and $Mosaic$, $Mosaic$ and \textit{Core-shell} connect with each other, but no stationary solution in \textit{Core-shell} class can connect to a lower-index saddle point in \textit{Tai-chi} class. 
These four classes play key role in connecting the high-index (index $>6$) saddle points to the transition states ($1$-saddles) and minima. 
For example, we observe the \textit{Flower} class can be connected from the well defined 7-saddles $Rhombus~Y$ and $Flower~8$ and then connect to the lower-index saddle points $Flower~4$,  $Dendritic~3N2$ and $Layer~4N$.
We can observe the \textit{Mosaic} class can be connected from almost all 7-saddles and connect to almost all 1-saddles. %except of $Rhombus~Y$, except of \textit{Core-shell~2B2}. 
%While, \textit{Core-shell} class can only be connected down from 7-saddle ($Triangle~2$) and then connect to three 1-saddles ($Layer~4N$, $Dendritic~3N3$, and \textit{Core-shell~2B2}).
%At last, the \textit{Tai-chi} class can be connected down from 7-saddles ($Triangle~2$, $Tai-chi~N$) and then connect to 1-saddles ($Layer~4N$, $Dendritic~3N3$).
Here, we treat each class as a whole and omit the detailed connections between the members of each class and the $1$ (or $7$)-saddles. 

\begin{figure}[hbt]
\centering
\includegraphics[width=0.48\textwidth,clip==]{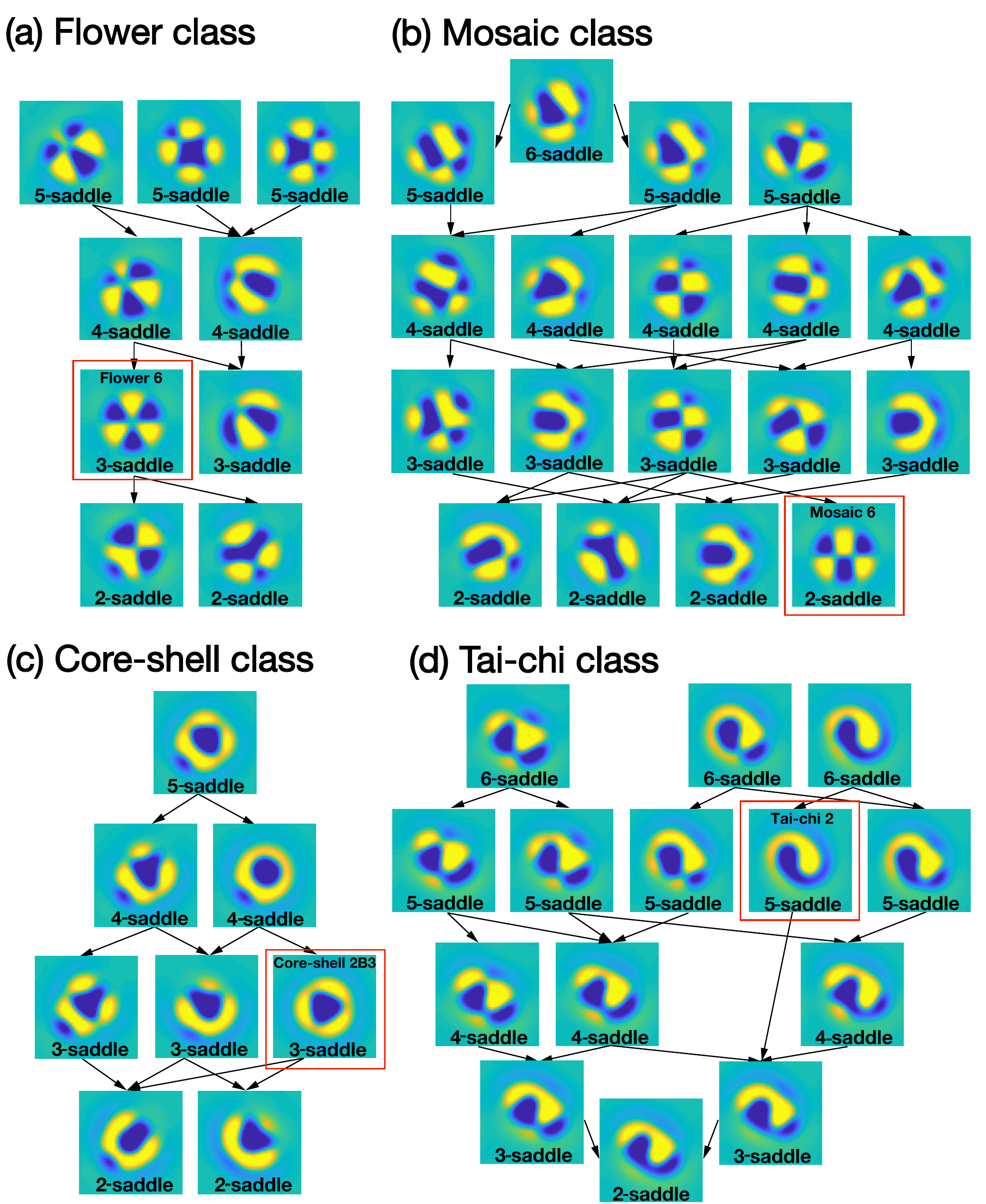}
\caption{Solution landscapes of the \textit{Flower} class (a), \textit{Mosaic} class (b), \textit{Core-shell} class (c) and \textit{Tai-chi} class (d). The typical solutions in Fig. \ref{sl_b14}: \textit{Flower~6}, \textit{Mosaic~6}, \textit{Core-shell~2B3} and \textit{Tai-chi~2} are marked with red box.
}\label{class_b14}
\end{figure}

%% Fig 6: solution landscape for Tai-chi; Flower class, Mosaic and Circle:
Fig.\ref{class_b14} shows four solution landscapes of \textit{Flower} class, \textit{Mosaic} class, \textit{Core-shell} class, and \textit{Tai-chi} class.
In each class, different solutions have subtle differences. 
%Since most morphologies in Fig. \ref{class_b14} are non-typical and less important, we omit their labels.
The \textit{Flower} class, including a typical state $Flower~6$ and many axisymmetric structures, was shown in Fig. \ref{class_b14}(a). 
%We can also observe the typical flowers, from $7$-saddle $Flower~8$ with 8 petals, $3$-saddle $Flower~6$ with 6 petals, to $1$-saddle $Flower~4$ with 4 petals.
The \textit{Mosaic} class with the richest solutions is shown in Fig. \ref{class_b14}(b). For instance, $2$-saddle $Mosaic~6$ is the typical mosaic, which looks like a floor tile with three yellow blocks alternating with three blue blocks in the elliptic confinement.
The \textit{Core-shell} class can connect to the circle phase $Circle~2B$ (minimum), shown in Fig. \ref{class_b14}(c). The typical core-shell is 3-saddle \textit{Core-shell~2B3}.
At last, we also observe the beautiful \textit{Tai-chi} class in Fig. \ref{class_b14}(d). The typical tai-chi phase is 5-saddle \textit{Tai-chi~2} with long tails and finally connects to a 2-saddle tai-chi with short tails.
%including the solutions with interior blue block surrounded by yellow block, is shown in Fig. \ref{class_b14}(c), which 
%The effect of domain size on phase structure is then well demonstrated by these stationary solutions with rich and fine internal structures, providing good guidance for diblock-copolymer experiments in 2D confinement. %region

\begin{figure}[hbt]
\centering
\includegraphics[width=0.4\textwidth,clip==]{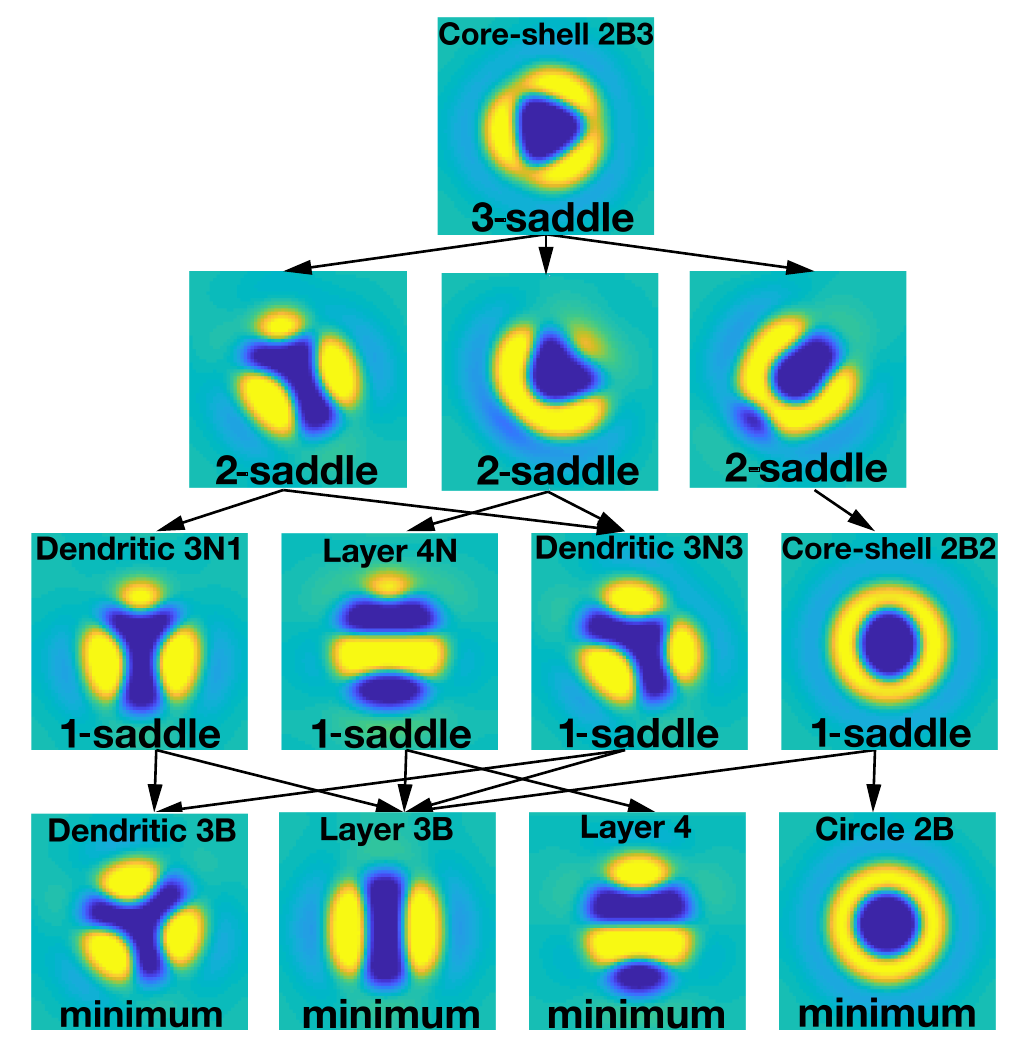}
\caption{Solution landscape starting from \textit{Core-shell~2B3} solution. All stable states including \textit{Dendritic~3B}, \textit{Layer~3B}, \textit{Layer~4} and \textit{Circle~2B} are connected by the index-3 \textit{Core-shell~2B3} solution.}\label{mountain}
\end{figure}

% Figure 7: transition between stable solutions
From the solution landscape in Fig. \ref{sl_b14}, there exist multiple transition pathways between the stable states (\textit{Dendritic~3B}, \textit{Layer~3B}, \textit{Layer~4} and \textit{Circle~2B}). For example, there are two transition pathways between \textit{Dendritic~3B} and \textit{Layer~3B}: $Dendritic~3B \rightarrow Dendritic~3N1 \rightarrow Layer~3B$ and $Dendritic~3B \rightarrow Dendritic~3N3 \rightarrow Layer~3B$.
%The transition pathway between \textit{Dendritic~3} and \textit{Layer~4} is $Dendritic~3 \rightarrow Dendritic~3N2 \rightarrow Layer~4$;
%The transition pathway between \textit{Layer~3B} and \textit{Layer~4}   is $Layer~3B \rightarrow Layer~4N  \rightarrow Layer~4$;
%The transition pathway between \textit{Layer~3B} and \textit{Circle~2B} is $Layer~3B \rightarrow \textit{Core-shell~2B2} \rightarrow Circle~2B$.
However, not all pairs of stable states can be connected by a single transition state. For example, \textit{Circle~2B} cannot be directly connected to \textit{Dendritic~3B} or \textit{Layer~4\textit}. Thus, the transition pathways between them need multiple transition states.
More specifically, the transition pathway between \textit{Circle 2B} and \textit{Dendritic~3B} can be $\textit{Circle~2B} \rightarrow \textit{Core-shell~2B2} \rightarrow \textit{Lay~3B} \rightarrow \textit{Dentritic~3N3} \rightarrow \textit{Dendritic~3B}$.
Fig. \ref{mountain} shows the 3-saddle \textit{Core-shell~2B3} is the stationary solution in the intersection of the smallest closures of all four stable states.
The dynamical pathways from \textit{Core-shell~2B3} can be constructed to connect every stable state passing through one 2-saddle and one 1-saddle. Our numerical results highlight the differences between transition pathways mediated by multiple transition states (1-saddles) and dynamical pathways mediated by single high-index saddle point.

\subsection{Symmetry breaking}
Finally, we investigate the symmetry-breaking phenomena shown in the solution landscapes with different preference intensity and domain size.
In view of the previous results, symmetric (reflectional and rotational) properties are quite useful to classify the morphologies for the equal preference as in Fig. \ref{sl_b0} and Fig. \ref{sl_b14}. 
%Similar symmetric morphologies also exist in the case of selective preference in Fig. \ref{sl_b1}.
The solutions in $D^k$ coincide by itself when rotating by angles $\frac{2\pi j}{k}, j = 1,...,k-1$, or reflecting about the symmetry axes.
For instance, the highest-index homogeneous phase has $D^\infty$ symmetry (any rotation and reflection is allowed).
% and other symmetric phases (in $D^k, k\geqslant 2$) are classified and shown in Fig. \ref{symmetry_1}.

\begin{figure}[hbt]
\centering
\includegraphics[width=0.48\textwidth,clip==]{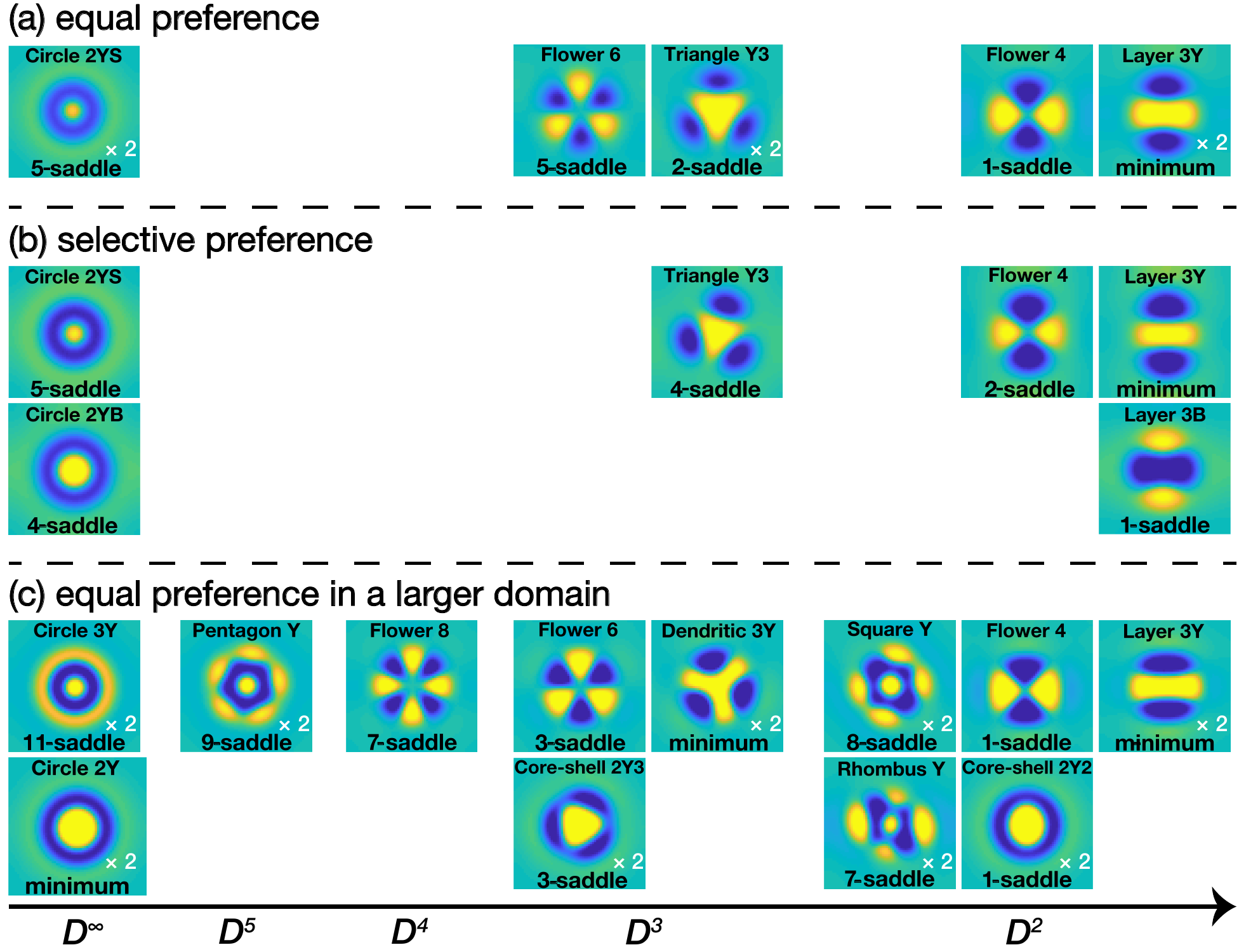}
\caption{The comparison of symmetric properties of the solutions in the solution andscapes with (a) equal preference in Fig. \ref{sl_b0}, (b) selective preference in Fig. \ref{sl_b1}, and (c) equal preference in a larger domain in Fig. \ref{sl_b14}. The notation "$\times 2$" means there exists a pair of phases, e.g. $Circle~2YS$ and its counterpart $Circle~2BS$.} \label{symmetry_1}
\end{figure}

In Fig. \ref{symmetry_1}, one can observe the symmetry-breaking process from circular-shape states that has $D^\infty$ to the lower symmetric solutions with some $D^k (2\leq k< \infty)$ symmetry, e.g. (a) \textit{Circle~2YS} ($D^{\infty}$) $\Longrightarrow$ \textit{Flower~6} and \textit{Triangle~Y3} ($D^3$) $\Longrightarrow$ \textit{Flower~4} and \textit{Layer~3Y} ($D^2$); (b) \textit{Circle~2YS} and \textit{Circle~2YB} ($D^{\infty}$) $\Longrightarrow$ \textit{Triangle~Y3} ($D^3$)$\Longrightarrow$ \textit{Flower~4}, \textit{Layer~3B}, and \textit{Layer~3Y} ($D^2$); and (c) \textit{Circle~3Y} ($D^{\infty}$) $\Longrightarrow$ \textit{Pentagon~Y} ($D^5$) $\Longrightarrow$ \textit{Flower~8} ($D^4$) $\Longrightarrow$ \textit{Flower~6}, \textit{Core-shell~2Y3}, and \textit{Dendritic~3Y} ($D^3$) $\Longrightarrow$ \textit{Square~Y}, \textit{Rhombus~Y}, \textit{Flower~4}, \textit{Core-shell~2Y2}, and \textit{Layer~3Y}($D^2$).

Fig. \ref{symmetry_1} also shows the effect of the preference intensity and the domain size on the symmetric property.
%the number of stationary solutions with symmetry $D^k$ ($k\geqslant 2$) is effected by the preference intensity and the domain size. 
In the case of equal preference, there are $8$ symmetric solutions in Fig. \ref{symmetry_1}(a). 
While, as $b_1$ is changed from $0$ to $0.1$, the selective preference breaks the symmetry and the number of symmetric solutions is reduced to $6$ in Fig. \ref{symmetry_1}(b). As stated before, \textit{Flower~6} and  \textit{Triangle~B3} disappeared. 
For the case of equal preference in a larger domain (Fig. \ref{symmetry_1}(c)), the number of symmetric solutions increases to $21$. This is because the larger domain can accommodate more copolymer molecules and multiple finer structures, e.g., the novel phases \textit{Pentagon~Y}, \textit{Flower~8}, \textit{Square~Y} and \textit{Rhombus~Y}.

We also note that the highest symmetry $D^k$ increases from $D^3$ to $D^5$ when the domain size $\lambda$  is changed from $1$ to $1.4$, except of $D^\infty$ . Thus, we expect that, for the stationary solutions of polygonal shapes (except of the homogeneous and circle phases), the upper bound of $k$ in symmetry group $D^k$ will continuously increase with the increase of the domain size in the case of equal preference. 
%While since the domain size is fixed and finite, there is a limitation of upper bound of polygonal shape (except of $D^{\infty}$) such that we cannot make arbitrary fine pattern with some finite fixed characteristic domain size. But we can expect more symmetric solutions with high symmetric property by increasing the domain size.% or shrinking the polymer length. 

\section{Conclusions and Discussions}
In this work, we studied the solution landscapes of the diblock copolymers and homopolymers in a 2D confinement with the  DCH model. The PSD method is developed to efficiently compute the high-index saddle points with mass conservation and construct the solution landscape of the DCH free-energy functional by coupling with downward/upward search algorithms.

We systematically constructed the solution landscapes by varying the preference intensity and the domain size. 
In the case of equal preference, branching from the homogeneous phase, the solution landscape was obtained, including the \textit{Circle} solution, \textit{Flower}-type solutions, \textit{Triangles}-type solutions and \textit{Layer}-type solutions. 
Furthermore, the solution landscape reveals the informative transition pathways between stable/metastable phases.
% and how the two blocks evolve and exchange their positions between a pair of the metastable solutions $Layer~3Y$ and $Layer~3B$ is answered.
While, with selective preference, the solution landscape loses nearly half of the stationary solutions compared with the one of the equal preference. 
We can observe the symmetry-breaking phenomenon and the surface has more preference for the blue blocks, which starts to surround the yellow blocks, with the increase of the preference intensity $b_1$. These results are consistent to the work by E. Avalos et al. \cite{Nishiura2016} and the solution landscape provides good guideline for experiments. 
%cases 3 larger domain
In the case of equal preference in a larger domain, the solution landscape shows more novel and interesting stationary solutions. In particular, we can classify the solutions as \textit{Flower} class, \textit{Mosaic} class, \textit{Core-shell} class, and \textit{Tai-chi} class. 
The relationships between different stable/metstable states are shown by the dynamical pathways connected by a single high-index saddle point (\textit{Core-shell~2B3}).  % case 4: symmetry
We further demonstrate that the symmetry-breaking phenomena largely exist in the solution landscapes from high-index saddle points to local minima. 
The number of stationary solutions with symmetry $D^k$ and the upper bound of $k$ increase as the domain size increases.% symmetry group,in $D^k$
%We note that we only consider the frustrated phases under confinement, namely having closed surface in the domain and having micro-phase separations. 
%Actually, there exists some non-confined stationary solutions, e.g. the rod-like surface along the diagonal of the square domain. 
%The pure-color circular solution can also be observed, while no micro-phase separation in the macro-phase domain(in the circular).

The numerical results of solution landscapes of the DCH model propose several follow-up questions.
First, the extension of the solution landscape from 2D to 3D is a fertile ground. The richer and more complex solutions are expected in the solution landscape in the 3D confinement, such as the multipod solution and twisted solution.~\cite{Nishiura2016} 
%With that, we may answer highly-demanded questions that occur in 3D space, for instance, how the solution landscape changes from lamellar phase to onion phase, as was discussed by E. Avalos et al. ~\cite{Nishiura2018}
On the other hand, the viewpoint of symmetry becomes more important in 3D case and more studies will be proceeded in the subsequent work.
The framework of solution landscape is believed to be a promising approach to find new exotic morphologies with the tight control of the initial conditions (saddle point and associated unstable directions), which overcomes the difficulty of tuning initial guesses to search stationary solutions needed in most numerical methods.
To deal with the mass-conservation constraint, the PSD method is developed using the typical $L^{2}$ inner product.  
We note that the study of SD can also be extended to incorporate the use of different inner products for defining the dynamic systems. One way is to apply the Cahn-Hilliard dynamics using the $H^{-1}$ inner product to avoid imposing the additional conservation constraint.\cite{zhang2014finding} It is interesting to develop the high-index saddle dynamics using the $H^{-1}$ inner product to compute any-index saddle points with mass-conservation constraint in the future work.

{\bf Acknowledgment}
This work was supported by the National Natural Science Foundation of China No. 12050002, 11971002, 11861130351. Y. Nishiura gratefully acknowledges the support  by JSPS KAKENHI Grant-in-aid No. 20K20341. 
Y. Han gratefully acknowledges the support from a Royal Society Newton International Fellowship.

\bibliographystyle{rsc}
\bibliography{references.bib}{}
\end{document}